# Particle Swarm Optimization for Mobility Load Balancing SON in LTE Networks

Zwi Altman(1), Soumaya Sallem(2), Ridha Nasri(1), Berna Sayrac(1) and Maurice Clerc(3)
(1)Orange Labs, 38 rue du Général Leclerc, 92794 Issy les Moulineaux, France
(2)CEA, Saclay, PC172 - 91191 Gif-Sur-Yvette, France
(3)204, route de la Nerulaz, 74570 Groisy, France

***Abstract*—This paper presents a self-optimizing solution for Mobility Load Balancing (MLB). The MLB-SON is performed in two phases. In the first, a MLB controller is designed using Multi-Objective Particle Swarm Optimization (MO-PSO) which incorporates *a priori* expert knowledge to considerably reduce the search space and optimization time. The dynamicity of the optimization phase is addressed. In the second phase, the controller is pushed into the base stations to implement the MLB SON. The method is applied to dynamically adapt Handover Margin parameters of a large scale LTE network in order to balance traffic of the network eNodeBs. Numerical results illustrate the benefits of the proposed solution.**

## I. Introduction

The Self-Organizing Network (SON) technology has been receiving a growing interest for two main reasons. The Radio Access Networks (RAN) landscape is becoming increasingly complex and heterogeneous with co-existing and co-operating technologies. In parallel, network operators experience growing pressure to reduce operational expenditure (OPEX). The SON technology has been introduced as a means to manage complexity, to reduce cost of operation, and to enhance performance and profitability of the network. Self-organizing networks aim at autonomously configure newly deployed network nodes (self-configuration), tune parameters to improve Key Performance Indicators (KPIs) (self-optimization) and perform diagnostic and reparation of faulty network nodes (self-healing) [1].

Self-optimization mechanisms are control loops that can be deployed in the management- or the control-plane. Control plane SON can be implemented in the eNodeBs (eNB) and benefit from high reactivity to traffic variations. For this reason, these SON solutions are sometime denoted *on-line SON*. On the other hand, management-plane (or *off-line*) SON solutions are deployed in the Network Management System (NMS), and are interfaced to the operator Operation and Maintenance Center (OMC) by means of APIs. Off-line SON solutions benefit from abundant data (metrics and KPIs) and computational means necessary for processing and running powerful optimization methods. The main drawback of the centralized approach is related to the long time scale that is typically used, in the order of an hour and more.

Mobility Load Balancing (MLB) aims at balancing network load by adapting mobility parameters. MLB has been mainly studied in the context of control-plane SON [4-6], and only a few contributions have addressed the management-plane MLB problem [7]. The aim of this paper is to study a hybrid approach for MLB SON that is performed in two phases. In the first, expert knowledge is combined with the Multi-Objective Particle Swarm Optimization (MO-PSO). The former allows to derive a parameterized form of the SON function (or the controller) which is then optimized by the MO-PSO in the management-plane. In the second phase, the controller is pushed into the control-plane, e.g. within eNBs, for on-line implementation. The advantage of such a solution is that it benefits from both computation resources available at the management-plane, and from high reactivity when implemented in the control-plane. It is noted that a two phase approach for self-optimization load balancing has been considered in the framework of Fuzzy Q-Learning (FQL) [8-9]. This approach can be viewed as a fully control-plane approach, where both learning and control (exploration and exploitation in the learning nomenclature) are performed on-line. It is noted that a hybrid learning approach could also be implemented. The utilization of expert (or *a priori*) knowledge has also an equivalence in the Reinforcement Learning (RL), namely the definition of a parameterized form of a policy which is then optimized during the learning process (e.g. Policy Gradient RL [10]).

The paper is organized as follows: Section II introduces the system model. Section III presents the optimization approach for the static and dynamic cases, including the formulation of the MO-PSO, and the incorporation of expert knowledge to the solution. Simulation results for the MLB SON are presented in Section IV followed by concluding remarks in Section V.

## II. SYSTEM MODEL

This section describes the LTE interference and Signal-to-Interference-plus-Noise Ratio (SINR) model used in this work. Similar modelling has been used in other contributions (see for example [5]).

### A. *Interference and SINR model*

Let $I_{k,m}$ denote the average downlink interference perceived by a mobile $m$ connected to eNB $k$. The interference is generated by neighbouring eNBs utilizing the same frequency. The probability of eNB $i$ to transmit on the same sub-carrier utilised by eNB $k$ is equal to the load of eNB $i$, $L_i$, defined as the ratio of the number of used Physical Resource Blocks (PRBs) to the total number of available PRBs. $I_{k,m}$ can then be written as [5]:

$$I_{k,m} = \sum_{i \neq k} \Lambda(k,i) L_i \frac{P_i G_{i,m}}{Q_{i,m}} \tag{1}$$

where $\Lambda$ stands for the interference matrix and the matrix element $\Lambda(k,i)$ equals one if cells $k$ and $i$ utilize the same frequency band and zero otherwise; $P_i$ denotes the downlink transmitted power per sub-carrier of eNB $i$; $G_{i,m}$ and $Q_{i,m}$ are the antenna gain and the path loss between eNB $i$ and the mobile $m$, respectively. It is noted that for data services, when there is a single mobile in a cell, certain technological implementations allow to allocate the entire eNB frequency bandwidth (i.e. all available PRBs) to the mobile. In this case $P_i$ is the maximum eNB power and $L_i$ - the proportion of transmission time. The SINR of mobile $m$ attached to eNB $k$ is derived from (1) as follows [5]:

$$SINR_{k,m} = \frac{P_k G_{k,m}}{Q_{k,m}\left(I_{k,m} + N_{th}\right)} \tag{2}$$

$N_{th}$ being the thermal noise per sub-carrier. The adaptive modulation and coding scheme is determined from the SINR value through the perceived BLoc Error Rate (BLER). In the network simulator, the throughput per PRB for each user is determined as a function of the SINR via link level curves. It is assumed here that the user physical throughput equals $N_m$ times the throughput per PRB, where $N_m$ is the number of PRBs allocated to user $m$.

### B. *Handover model*

3GPP standard has specified hard handover for mobility in the LTE network however the implementation of the handover algorithm is not specified in the standard. A GSM-like hard handover algorithm is proposed. A handover condition on the trasmitted pilot power is defined as:

$$P_i - P_k \geq HM(k,i) + Hysteresis \tag{3}$$

and $i$ (in dB) respectively and $HM(k,i)$ is the Handover Margin (HM) between eNB $k$ and $i$. Hysteresis is a counter-measure that prevents ping-pong effect by providing a baseline handover threshold against power fluctuations due to channel variations. For the sake of simplicity, *Hysteresis* is set to zero and the baseline threshold value is integrated into $HM(k,i)$. In addition to (3), the following two conditions should be verified:

- The received power from the target eNB should be higher than a predefined threshold, and
- Enough resources should be available in the target eNB.

## III. HANDOVER OPTIMIZATION

### A. *Expert knowledge*

Expert knowledge refers to *a priori* knowledge on the optimization problem. It gives rough information (or tendency) on the type of parameter modification that will improve the system performance in different states of the system. Expert knowledge can be used to guide the optimization process and to reduce the search space. It is typically given in the form of a set of rules that relates qualitatively the system state to the parameter $x$. The system state is defined by a vector $\vec{u} = (u_1, \cdots, u_m)$ of system indicators such as eNB load or interference, and often does not give direct information on the user perceived QoS.

As an example, assume that *HM*(*i*,*j*) depends on the load of station *i*, $L_i$, and that of its neighbor *j*, $L_j$. Denote by $HM_0$ the planning (or default) value used in the network planning process. Then the expert knowledge can be given in the form of the following four rules:

(*i*) If $\langle L_i \text{ is Low}\rangle$ and $\langle L_j \text{ is Low}\rangle$ then $\langle \text{set } HM(i,j) \text{ to } HM_0\rangle$

(*ii*) If $\langle L_i \text{ is High}\rangle$ and $\langle L_j \text{ is High}\rangle$ then $\langle \text{set } HM(i,j) \text{ to } HM_0\rangle$

(*iii*) If $\langle L_i \text{ is Low}\rangle$ and $\langle L_j \text{ is High}\rangle$ then $\langle \text{set } HM(i,j) \text{ to high value}\rangle$

(*iv*) If $\langle L_i \text{ is High}\rangle$ and $\langle L_j \text{ is Low}\rangle$ then $\langle \text{set } HM(i,j) \text{ to low value}\rangle$

Rules (*i*) and (*ii*) prevent unnecessary handovers (ping-pong effects). Rule (*iii*) aims at helping the loaded eNB *j* by delaying handovers from eNB *i*. Rule (*iv*) aims at alleviating the loaded eNB *i* by advancing handovers towards eNB *j*.

In the general case, the parameter *x* is a function of the vector $\vec{u}$. For example the parameter *x* stands for the Handover Margin (HM), and the vector $\vec{u}$ – for the load vector ($L_i$,$L_j$) (Figure 1). We write *x* as $x = surf(\vec{u})$ where *surf* stands for a multi-dimensional surface (i.e. the control function). Experience in HM optimization shows that the parameter function *x* varies smoothly with $\vec{u}$. The expert knowledge given by a set of rules is used to guess a simple form for the function *surf*. Our aim is to find a parametric representation of *surf*, namely to write it as a function of a parameter vector $\vec{p}$, $\vec{p} = (p_1, \cdots p_m)$ with a few elements:

$$x = surf(\vec{u}; \vec{p}) \tag{4}$$

Hence the form of the function *surf* is fully defined by the vector $\vec{p}$ that is determined via an optimization process. The function *surf* is denoted hereafter as the *parameterization surface*.

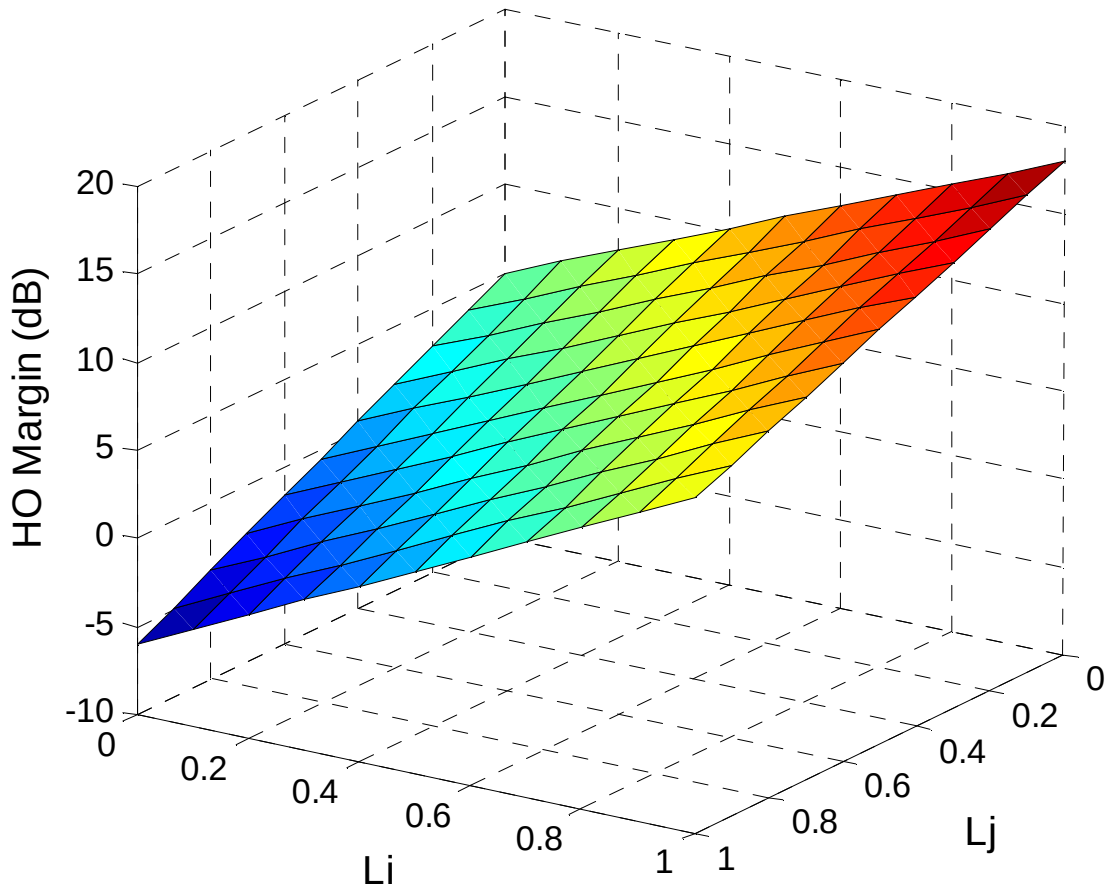

**Fig. 1.** Polynomial parameterization surface of HM in the form of a linear interpolation of the extreme points: *HM*(0,0), *HM*(1,0), *HM*(0,1) and *HM*(1,1), as a function of the loads of two neighboring eNBs.

## B. *Multi-objective optimization model*

Denote by $\vec{F}(x) = (f_1(x), \ldots, f_n(x))$ the objective function to be optimized where $f_i(x)$ represents a KPI which depends on the parameter *x*. It is recalled that the parameter *x* has a parametric representation given by (4), hence $\vec{F}$ can be written as a function of $\vec{p}$. The element $p_i$ is defined within an interval [ $p_{i_{min}}, p_{i_{max}}$ ]. The optimization problem is a constraint multi-objective optimization problem and is written as follows:

$$
\begin{aligned}
&\textit{maximize :}\quad \vec{F}(x)=\left(f_1(x), f_2(x),\ldots, f_n(x)\right)\\
&\textit{subject to :}\quad p_i \in \left[p_{i_{min}}, p_{i_{max}}\right];\quad i=1,\cdots,m
\end{aligned}
\tag{5}
$$

Pareto optimal solutions are sought. A solution is Pareto optimal if the objective vector cannot be improved in any dimension without degradation in another dimension. The solution $x$ dominates the solution $x'$ if and only if

$$
\begin{cases}
\forall\, i \in \{1,\ldots,n\}\ ,\ f_i(x) \geq f_i(x') \\
\qquad\qquad and \\
\exists\, j \in \{1,\ldots,n\} \mid f_j(x) > f_j(x')
\end{cases}
\tag{6}
$$

A solution is said to be non-dominated if there exists no solution that dominates it. The set of non-dominated solutions within the entire search space constitute the Pareto optimal front.

### *C. HM Parameterization Surfaces*

This section is devoted to the construction of two parameterization surfaces for the *HM* parameter. According to (4) we can write:

$$
HM(i,j) = surf_{HM}\left(\vec{L};\vec{p}\right) \tag{7}
$$

where $\vec{L}=\left(L_i, L_j\right)$, $L_i$ and $L_j$ being the load of eNB $i$ and of its neighbor $j$ respectively. According to the expert knowledge and the four rules presented in Section III.A (see Fig. 1), we define the function *surf* via a linear interpolation of the four extreme points: *HM*(0,0), *HM*(1,0), *HM*(0,1) and *HM*(1,1). It is noted that *HM*(0,0) and *HM*(1,1) can be chosen as the planning value $HM_0$ or as a value belonging to a small interval around $HM_0$:

$$
\begin{aligned}
surf_{HM}\left(\vec{L}\right) = {} & HM(0,0) \\
& + \left(HM(1,0) - HM(0,0)\right)\cdot L_i \\
& + \left(HM(0,1) - HM(0,0)\right)\cdot L_j \\
& + \left(HM(0,0) + HM(1,1) - HM(0,1) - HM(1,0)\right)\cdot L_i \cdot L_j
\end{aligned}
\tag{8}
$$

Eq. (8) has the form of a polynomial in the loads $L_i$ and $L_j$:

$$
surf_{HM}\left(\vec{L};\vec{p}\right) = b_0 + b_1 L_i + b_2 L_j + b_3 L_i L_j \tag{9}
$$

with $\vec{p}=\left(b_0, b_1, b_2, b_3\right)$. Figure 1 shows an example of the function $surf_{HM}\left(\vec{L};\vec{p}\right)$. The vector $\vec{p}$ is determined via an optimization process.

With the aim of further improving the solution for the HM, an exponential parameterization surface is investigated. This solution accentuates pushing (delaying) mobiles to make handovers for high (absolute value) load differences $|L_i\text{-}L_j|$ between neighboring cells. The exponential type of variation for $surf_{HM}$ as a function of the load difference $w$, $w = L_i - L_j$ , and is written as:

$$
surf_{HM}(w;\vec{p}) =
\begin{cases}
b \cdot e^{ln(a_1/b)w} & ;\quad w \geq 0 \\
2b - b\cdot e^{-ln(a_2/b)w} & ;\quad w < 0
\end{cases}
\tag{10}
$$

where $\vec{p}=\left(a_1, a_2, b\right)$. A 3D plot of $surf_{HM}$ in (10) with $b$=6, $a_1$=20 and $a_2$=20 is depicted in Figure 2. Results obtained using the optimized polynomial and exponential surfaces are presented in Section IV.

### *D. Particle Swarm Optimization*

The Particle Swarm Optimization (PSO) method is used here to optimize the parameter vector $\vec{p}$ that defines the parameterization surface. The PSO method is a robust technique belonging to the category of *Swarm Intelligence* methods which is inspired by the social behaviour of flocking organisms [2-3]. It utilizes a population of particles, each of which represents a solution, namely a parameter vector defining the parameterization surface. In the PSO notation, the position of a particle $i$, $\vec{p}_i$ stands for the parameter we seek to optimize (i.e. the parameter vector defining the parameterization surface in the present work). The particles probe regions in the solution space in a partially random way. The exploration of a particle is described in terms of a *velocity* $\vec{v}_i$, which is added to the current position to bring the particle to its next position. Hence the velocity stands for the update brought to a current solution. The velocity of a particle comprises three components which depend on its past best position, $\vec{q}_i$, on the past best position of its neighbours, $\vec{g}_i$, and on its own (current) velocity (Figure 2).

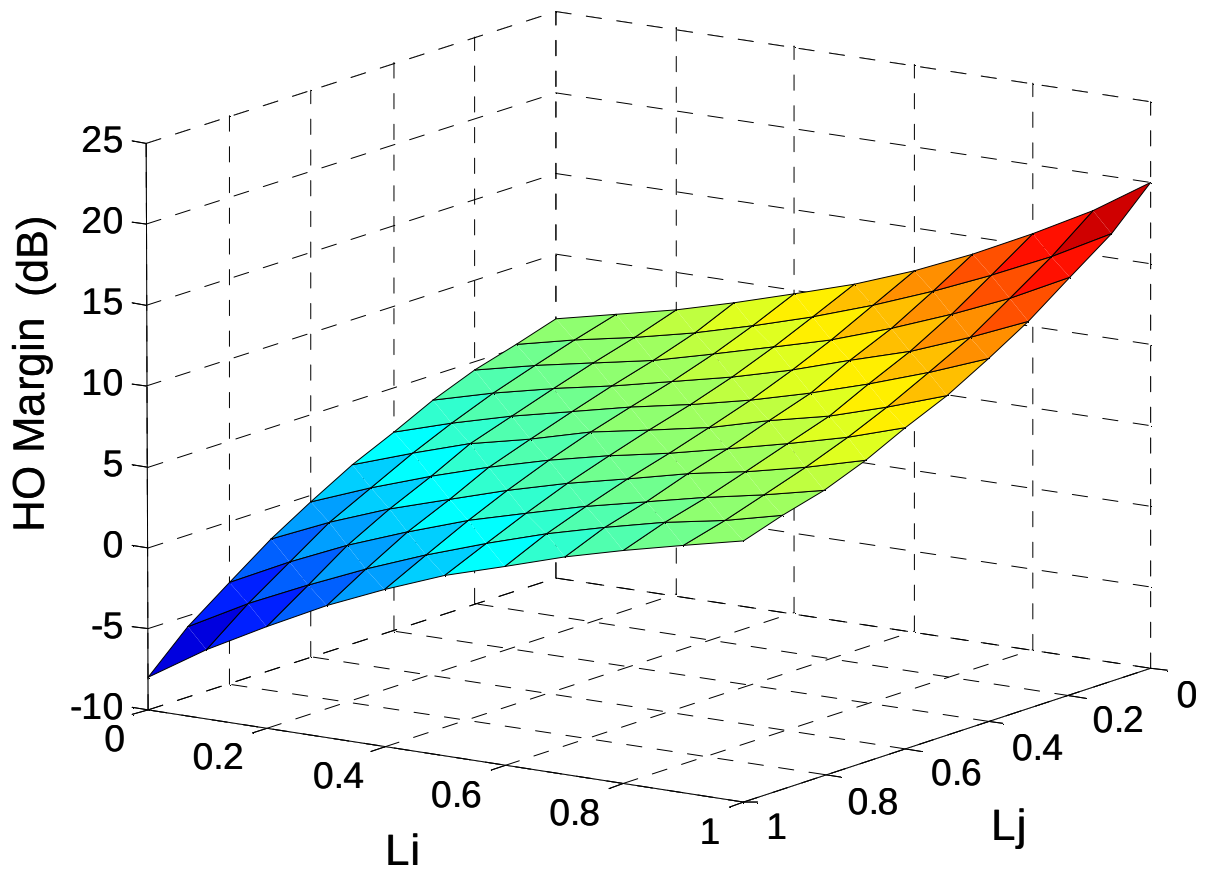

**Fig. 2.** Exponential parameterization surface for HM.

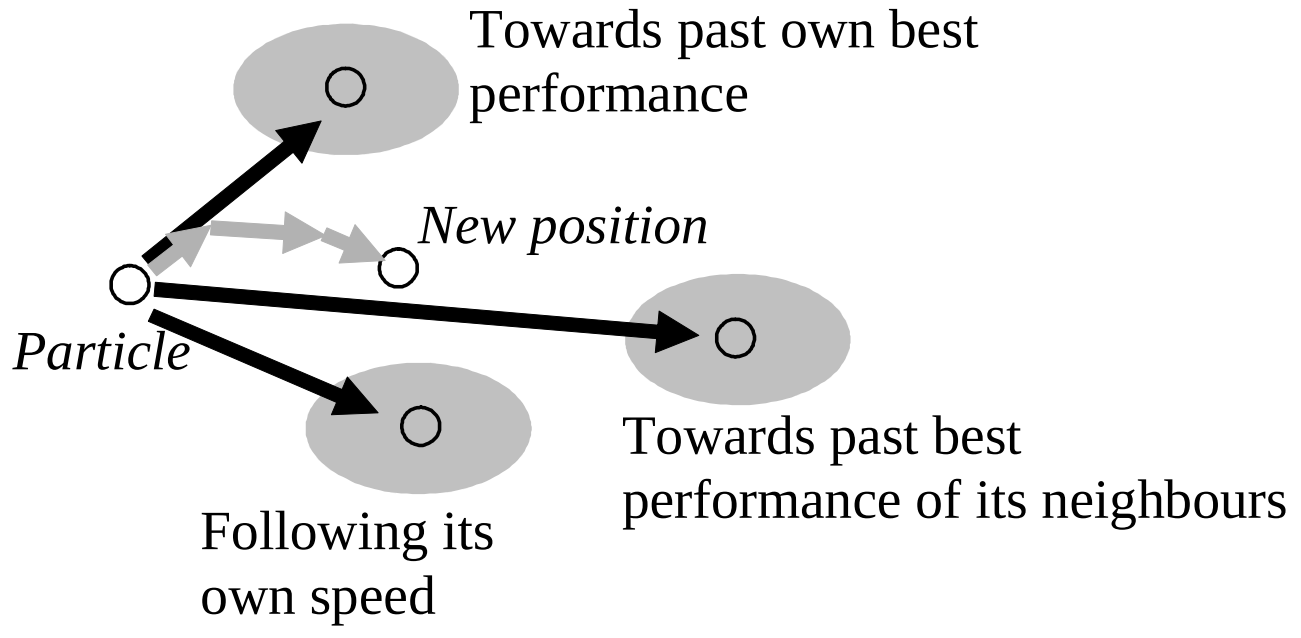

**Fig. 3.** Update of a particle position in the solution space.

The optimization process is carried out as an optimization loop. At each iteration, the particles evolve using the following update equation applied to each dimension (element) *d* of the particle:

$$\begin{cases} v_i^d \leftarrow c_1 v_i^d + c_{max}\, rand(0,1)\left(q_i^d - p_i^d\right) + c_{max}\, rand(0,1)\left(g_i^d - p_i^d\right) \\ p_i^d \leftarrow p_i^d + v_i^d \end{cases} \quad (11)$$

where *rand*(.) denotes the random function. It has been shown that $c_1$ and $c_{max}$ can be derived analytically from a single parameter $\varphi$ [11], which has been set to 4.14 here. The problem at hand is multi-objective, with two objectives to be optimized, $\vec{F}(\vec{p}_i) = (f_1(\vec{p}_i), f_2(\vec{p}_i))$: the network throughput and the access probability defined as the ratio between the number of successful network access attempts to the total number of network access attempts. For each new solution (particle) the network is simulated and $\vec{F}(\vec{p}_i)$ is evaluated. In this work, the swarm dimension, namely the population size equals 10. The set of neighbors for each particle comprises 3 different particles chosen at random (a particle cannot serve as its own neighbor). Hence the vector $(\vec{g}_k)_i$ comprises 3 particles.

The pseudo code for the multi-objective PSO algorithm is given below:

```
Initialization
For i=1 to Population Size
  Initialize randomly p_i, q_i, and list
  of neighbours of p_i : (g_k)_i, k ∈ {1,2,3};
End
Repeat for N_max iterations
For i=1 to Population Size
1. If F(p_i) ≻ F(q_i) then q_i = p_i
2. g_i = max_k (g_k)_i
3. For each dimension, update p_i^d using equation (11)
End
```

The term '$\succ$' determines the multi-objective characteristic of the algorithm. We say that the particle $\vec{p}_i$ is better than $\vec{p}_j$, i.e. $\vec{F}(\vec{p}_i) \succ \vec{F}(\vec{p}_j)$, if the following condition is satisfied

$$\begin{gathered}\vec{F}(\vec{p}_i) \succ \vec{F}(\vec{p}_j)\; IF\; \left(f_1(\vec{p}_i) \geq f_1(\vec{p}_j)\; AND\;\; f_2(\vec{p}_i) \geq f_2(\vec{p}_j)\right) AND \\ \left(f_1(\vec{p}_i) > f_1(\vec{p}_j)\;\; OR \;\; f_2(\vec{p}_i) > f_2(\vec{p}_j)\right)\end{gathered} \tag{12}$$

The condition (12) states that the solution $\vec{p}_i$ is better than $\vec{p}_j$ if at least one of its objectives, $f_1$ or $f_2$, is strictly bigger whereas for the other objective, the relation '$\geq$' holds. In step 2 of the PSO algorithm, the maximum is calculated with respect to the operator $\succ$ defined in (12). If more than one non-dominated solution is found in a neighbouring set, i.e. for which the relation (12) is not verified, the first non-dominated solution encountered is chosen. This last simplification allows to use the multi-objective version of the PSO algorithm written above, which is simpler than the ones described in [12].

*E. Dynamic optimization*

The optimization framework presented above can be directly extended to perform dynamic optimization of the HM parameter, $HM(i,j)$, written hereafter as $HM_{ij}$. In the dynamic optimization, $HM_{ij}$ becomes a function of time:

$$HM_{ij}(t) = surf_{HM}\left(\vec{L}(t)\,;\,\vec{p}\right) \tag{13}$$

and is updated every $\Delta_t$ seconds (set here to 5 sec.) according to the load values that are used to sample $surf_{HM}$. Each solution for the surface $surf_{HM}(\vec{L}(t)\,;\vec{p})$ is evaluated as in the static optimization case, during a simulation period of 2000 (simulator) seconds, required to achieve convergence of the optimization objectives.

It is noted that the small dimension of the search space considered here allows to use the simple version of the PSO described in Section III.D in a dynamic context. For a bigger search space, more sophisticated "dynamic PSO" algorithms may be necessary with slower convergence properties [13].

## IV. RESULTS

Consider a LTE network composed of 45 eNBs in a dense urban environment (see Figure 4). A semi-dynamic simulator is used to simulate the LTE network in the downlink. The simulator performs correlated Monte Carlo snapshots with a time resolution of a second to account for the time evolution of the network. FTP data traffic is considered. The principles of a semi-dynamic simulator are described in [14]. The Okumura-Hata propagation model is used for the 2GHz band. The path loss at a reference distance of 1 km and the path loss exponent are chosen as -128 dB and 3.76 respectively. Shadowing is modeled as a log-normal random variable with a 6 dB standard deviation. The spectrum efficiency depends on the SINR and on the Adaptive Modulation and Coding scheme used. The mapping between the SINR and the corresponding spectral efficiency is carried out using a quality table incorporated within the simulator. Fast fading is implicitly taken into account by the quality table.

Each eNB has 15 PRBs, corresponding to a 3 MHz bandwidth allocation. A frequency reuse factor of 3 is used. FTP calls are generated using a Poisson process of parameter $\lambda = 5$ arrivals per second. The arrivals are uniformly distributed within the network area. The non-uniform eNBs' positions results in highly non-uniform load distribution as can be seen from the histogram of Figure 6b. Each user is allocated between one and four PRBs to download a file of 10 Mbits. The duration of a communication depends on the allocated resources and on the user spectral efficiency (or throughput). The simulation parameters are summarized in Table I.

The HM parameterization surface is optimized using the multi-objective PSO algorithm described in Section III.D. Each particle (solution) corresponds to a distinct parameterization surface and is defined by the parameter vector $\vec{p}$ defined in Section III.C for the polynomial and the exponential surfaces. Two objectives are used to guide the PSO optimization: the total throughput of all eNBs in the network averaged over the simulation period, *Throughput*, and the probability of accessing the network, $P_{access}$. For each particle the network is simulated during a period of $T_{simu}$= 2000 (simulator) seconds to allow *Throughput* and $P_{access}$ indicators to converge. The PSO algorithm uses a population size of 10 particles and is repeated $N_{max} = 30$ iterations, namely 300 fitness (solution) evaluations are performed. Both static (eq. (7)) and dynamic (eq. (13)) optimization are performed. In the static optimization, the loads used to sample the optimization surface are considered fixed during the entire optimization process. These load values are the average loads calculated from a single simulation of the network with the planning HM value, $HM_0 = 6$ dB.

TABLE I. SIMULATION PARAMETERS.

| Parameter | Setting |
|---|---|
| System bandwidth | 3MHz |
| Frequency reuse scheme<br>Cell layout | 3<br>45 eNBs, sectorized |
| Inter-site distance | 1.5 to 2 km |
| PRB per eNB<br>PRB assigned per mobile<br>PRB transmit power | 15<br>1 to 4 (first-come first-serve basis)<br>32 dBm |
| Thermal noise density<br>Traffic type | -174 dBm/Hz<br>FTP |
| File size | 10 Mbits |
| Path Loss | 128 + 37.6 $\log_{10}$ (R), R in km |
| Shadowing standard deviation | 6 dB |
| Mobility<br>Speed | 40% of users are mobile<br>30 km/h |

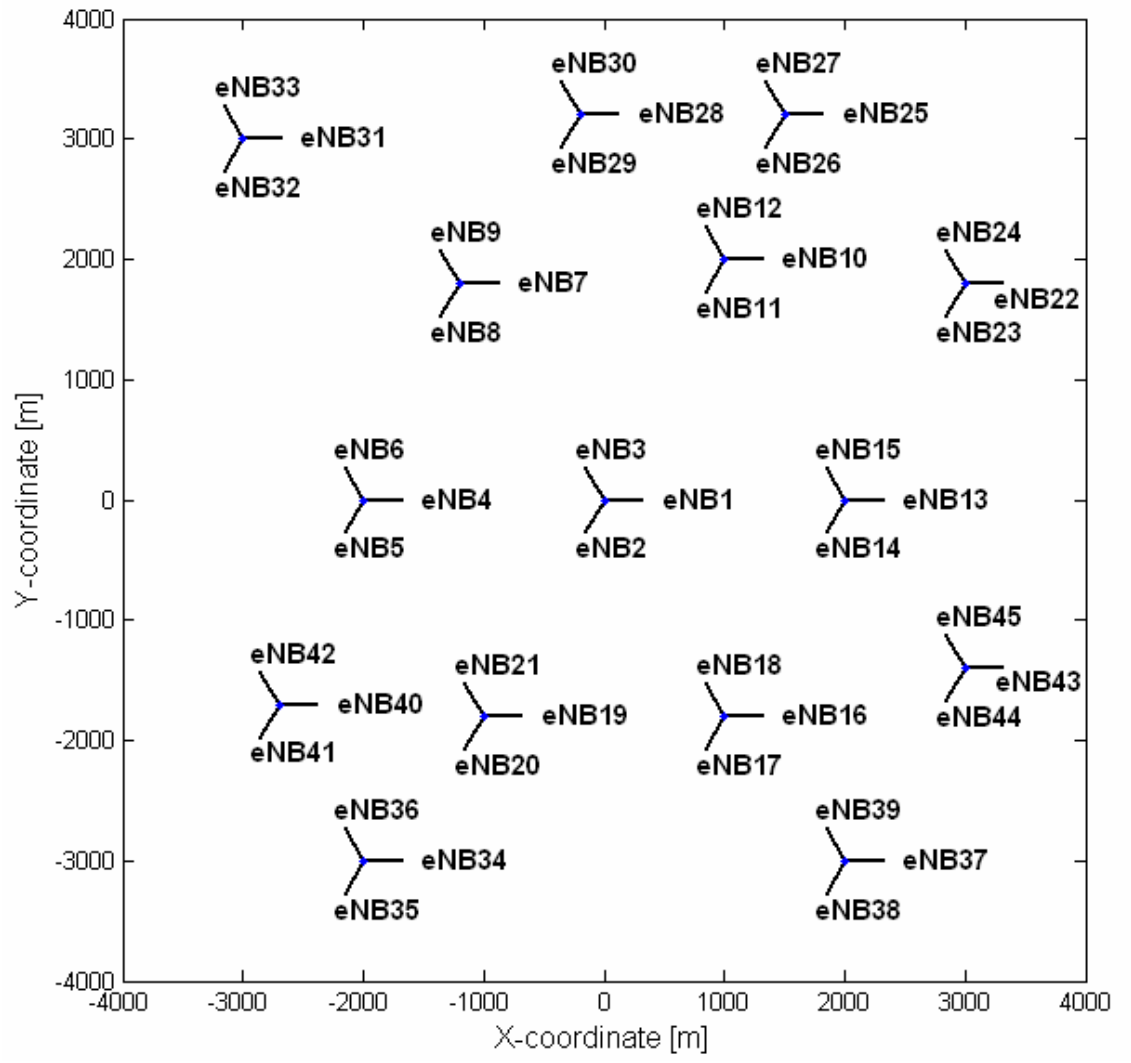

**Fig. 4.** A LTE network with 45 eNBs .

Figure 5 presents the MO-PSO results in the $P_{access}$–*Throughput* plane for both static and dynamic optimization. The solution corresponding to the planning $HM_0$ value (denoted hereafter as the planning solution) is plotted using a yellow circle. The first Pareto front in red triangles corresponds to the static optimization with the polynomial parameterization surface function (9). The results for the Pareto front using the exponential parameterization surface (10) are shown with blue squares in the Figure. The parameter $b$ in (10) is fixed to the planning value of 6 dB and the parameters $a_1$ and $a_2$ are optimized by the PSO. The Pareto-front for the exponential solution clearly dominates that of the polynomial solution. The results for the dynamic optimization using the exponential parameterization surface (13) are depicted in green diamond in Figure 6. Each eNB adapts the $HM_{ij}$ parameters every five seconds by sampling the exponential parameterization surface at the corresponding load values $L_i$ and $L_j$. One can see that dynamic optimization produces a better controller than the static optimization, and the performance gain using (13) improves the throughput and access probability by a few percent with respect to the planning solution.

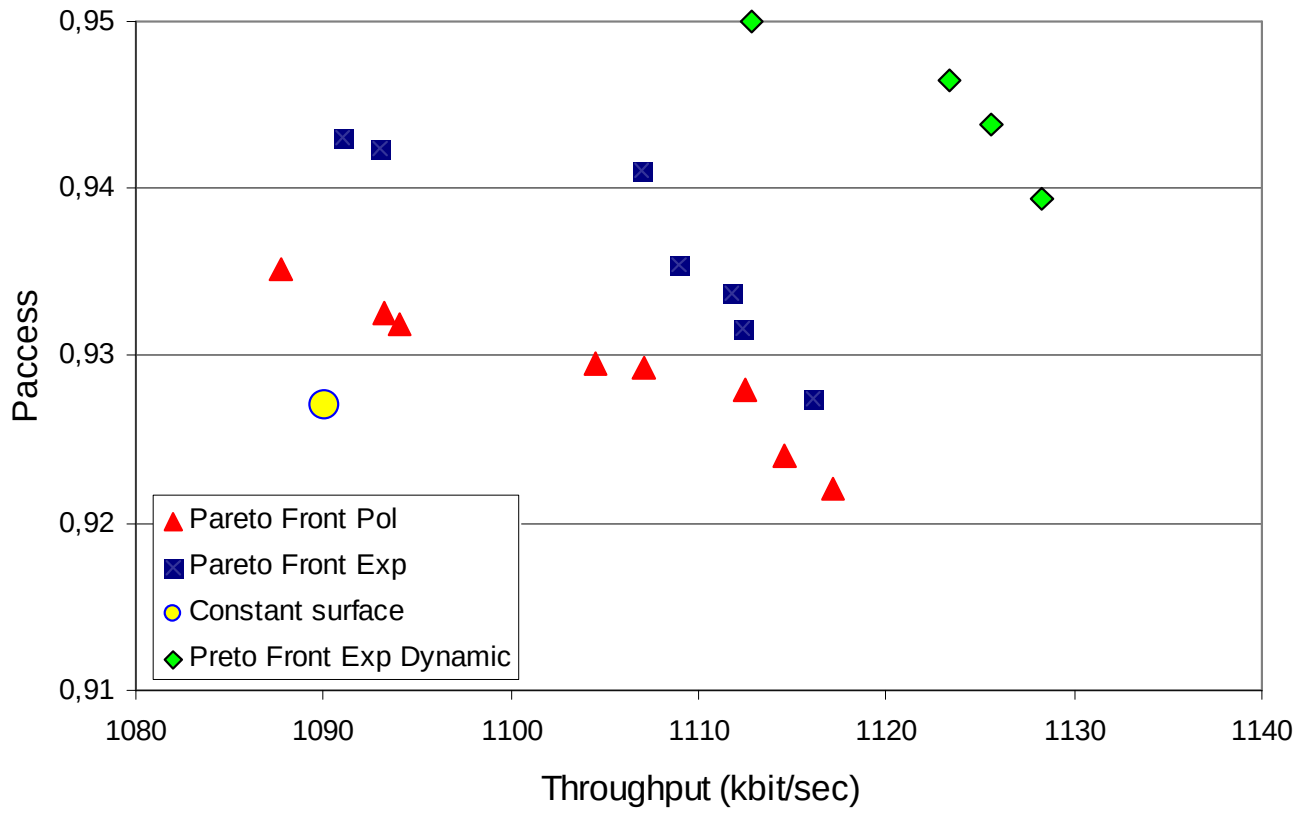

**Fig. 5.** Pareto-front solutions for constant optimization with polynomial (triangles) and exponential (squares) parameterization surfaces, and for the dynamic optimization using the exponential surface (diamonds) in the $P_{access}$–*Throughput* plane. The planning solution with $HM_0$ of 6 dB is plotted using a circle.

Denote the neighbor with which the eNB has the largest handover traffic exchange as the *best neighbour*. Figure 6a presents a histogram of $HM_{ij}$ of the eNBs and their corresponding best neighbours for a solution on the Pareto front for the polynomial parameterization surface in Figure 5. One can see that the optimization spreads the histogram to both low and high values. The optimized solution has many eNBs with low $HM_{ij}$ values (i.e. smaller than $HM_0$) allowing to advance the handovers. Several eNBs see their $HM_{ij}$ increase above 7 dB resulting in the delay of handovers towards loaded eNBs. The load histogram for the planning solution with $HM_0$=6 dB (white) and for the same optimized solution is depicted in Figure 6b. One can see that the MO-PSO reduces the number of eNBs with very high/low loads.

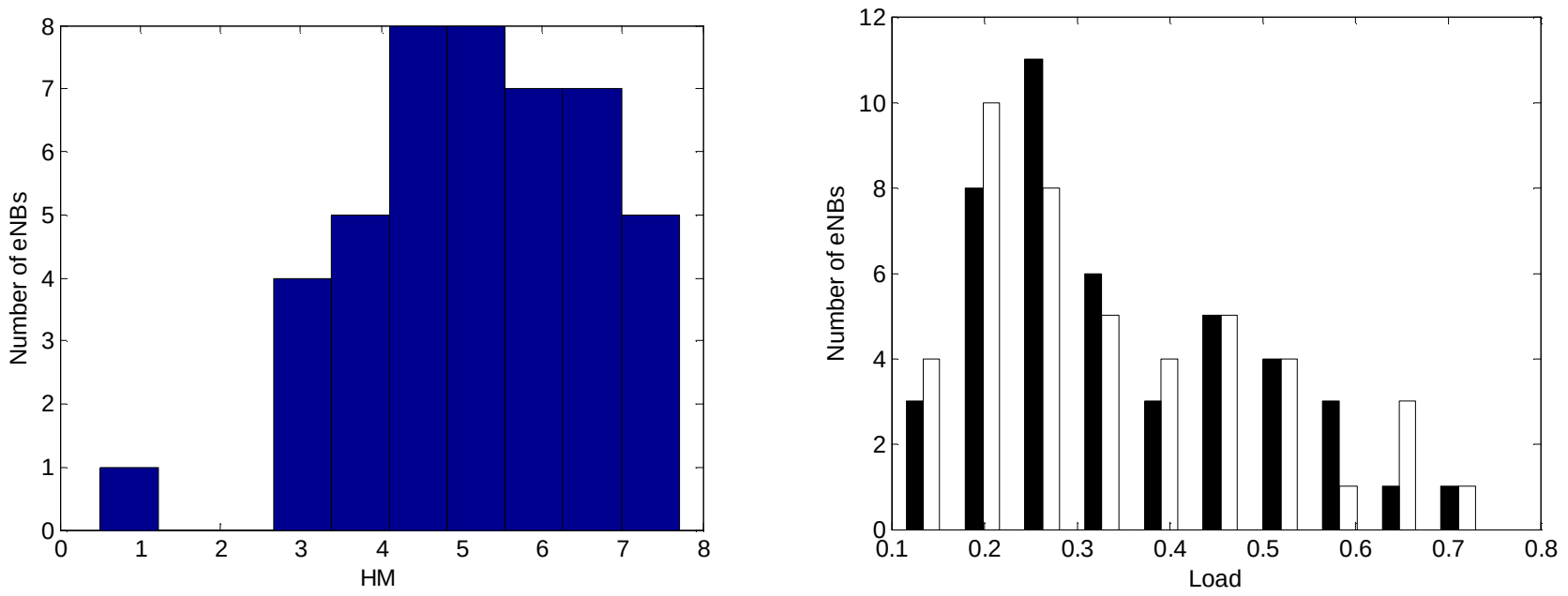

(a) (b)

**Fig. 6.** HM histogram for a solution on the Pareto front for the polynomial parameterization surface (a), and the corresponding load histogram for the planning (white) and an optimized (black) solutions (b).

## V. CONCLUSIONS

This paper has presented an efficient methodology for designing MLB SON controller. The method combines *a priori* expert knowledge with Multi-Objective Particle Swarm Optimization (MO-PSO), which allows to considerable reduce the search space and the computational time required for designing the MLB SON controller. The *a priory* knowledge provides the parameterized form of the controller which is optimized by the MO-PSO. It has been shown that dynamic optimization outperforms static optimization, namely produces better MLB-SON controller, which improve the throughput and access probability by a few percent with respect to the planning solution.

[1] 3GPP TR 36.902, "Evolved Universal Terrestrial Radio Access Network (E-UTRAN); Self configuration and self-optimization network use cases and solutions", (Release 8), Feb. 2008.

[2] J. Kennedy, R. Eberhart, and Y. Shi Y, Swarm Intelligence, Morgan Kaufmann, Academic Press, 2001.

[3] M. Clerc, Particle Swarm Optimization, ISTE (International Scientific and Technical Encyclopedia), 2006.

[4] J. Rodriguez, I. de la Bandera, P. Munoz, and R. Barco, “Load balancing in a realistic urban scenario for LTE networks,” IEEE 73rd Vehicular Technology Conference (VTC Spring 2011), May 2011.

[5] R. Nasri and Z. Altman, "Handover adaptation for dynamic load balancing in 3GPP long term evolution systems", 5th International Conf. on Advances in Mobile Computing & Multimedia (MoMM2007), Jakarta, Indonesia, Dec. 2007.

[6] R. Combes, Z. Altman, E. Altman, “Self-organization in wireless network: a flow-level perspective,” IEEE INFOCOM 2012, Orlando, US, March 2012.

[7] M. I. Tiwana, B. Sayrac, Z. Altman, T. Chahed, "Troubleshooting of 3G LTE mobility parameters using iterative statistical model refinement," *IFIP Wireless Days 2009*, Paris, Dec. 2009.

[8] P. Muñoz, R. Barco, I. de la Bandera, M. Toril, and S. Luna-Ramirez. Optimization of a fuzzy logic controller for handover-based load balancing”, IEEE 73rd Vehicular Technology Conference (VTC Spring 2011), May 2011.

[9] R. Nasri, Z. Altman et H. Dubreil, "Fuzzy Q-learning based autonomic management of macrodiversity algorithms in UMTS networks", Annales des Telecommunications, Sept.-Oct. 2006, pp.1119-1135.

[10] R. Combes, Z. Altman and E. Altman, “Self-Organizing relays: dimensioning, self-optimization and learning”, in IEEE Transactions on Network Management, TNSM, vol. 9, Dec. 2012 pp. 487-500.

[11] M. Clerc, and J. Kennedy, "The particle swarm: Explosion, stability, and convergence in a multi-dimensional complex space," IEEE Tran. Evol. Comput., vol. 6, Feb. 2002, pp. 58-73.

[12] M. Reyes-Sierra and C. A. C. Coello, “Multi-Objective Particle Swarm Optimizers: A Survey of the State-of-the- Art,” International Journal of Computational Intelligence Research, Vol. 2, No. 3, 2006, pp. 287-308.

[13] T. Blackwell. & J. Branke, G. R.Raidl (ed.), Multi-Swarm Optimization in dynamic environments applications of evolutionary computing, Springer, 2004, 3005 LNCS, pp. 488-599.

[14] A. Samhat et al, "Semi-dynamic simulator for large scale heterogeneous wireless networks", International Journal on Mobile Network Design and Innovation (IJMNDI), Vol. 1, N. 3-4, 2006, pp. 269-278.